%% file: paper.tex
\documentclass[noncommercial,copyright,creativecommons]{eptcs}

 \providecommand{\event}{QAPL 2010}
 \providecommand{\volume}{??}
 \providecommand{\anno}{2010}
 \providecommand{\firstpage}{1}
 \providecommand{\eid}{??}

\usepackage{inputenc}
\usepackage{pstricks}
\usepackage[dvips]{graphicx}

\pdfoutput=0

\newcommand{\failOrder}[3][\Gamma]{\ensuremath{#2\preceq_{#1}#3}}

\newcommand{\until}{{\bf\,U\,}}

\newcommand{\And}{\wedge}
\newcommand{\Or}{\vee}

\newcommand{\disjunction}{\bigvee}

\newcommand{\ftaFormula}{\ensuremath{\sum_{\Delta \in mcs}\prod_{\delta \in \Delta}}}

\newcommand{\true}{\ensuremath{true}}
\newcommand{\false}{\ensuremath{false}}

\newtheorem{definition}{{\bf Definition}}

\newcommand{\tempres}{\ensuremath{\delta t}}

\def\authorrunning{Matthias G\"udemann \& Frank  Ortmeier}
\def\titlerunning{Probabilistic Model-Based Safety Analysis}

\title{Probabilistic Model-Based Safety Analysis}

\author{Matthias G\"udemann\thanks{{\bf Acknowledgement:} Matthias
    G{\"udemann} is funded by the German Ministry of Education and Science
    (BMBF) within the ViERforES project (no. 01IM08003C)} \qquad \qquad  Frank  Ortmeier
\institute{Computer Systems in Engineering, Otto-von-Guericke University of Magdeburg}
\email{\{matthias.guedemann,frank.ortmeier\}@ovgu.de}}

\begin{document}

\maketitle

\begin{abstract}

  Model-based safety analysis approaches aim at finding critical failure
  combinations by analysis of models of the whole system (i.e. software,
  hardware, failure modes and environment). The advantage of these methods
  compared to traditional approaches is that the analysis of the whole system
  gives more precise results.

  Only few model-based approaches have been applied to answer quantitative
  questions in safety analysis, often limited to analysis of specific failure
  propagation models, limited types of failure modes or without system dynamics
  and behavior, as direct quantitative analysis is uses large amounts of
  computing resources.  New achievements in the domain of (probabilistic)
  model-checking now allow for overcoming this problem.

  This paper shows how functional models based on synchronous parallel
  semantics, which can be used for system design, implementation and qualitative
  safety analysis, can be directly re-used for (model-based) quantitative safety
  analysis. Accurate modeling of different types of probabilistic failure
  occurrence is shown as well as accurate interpretation of the results of the
  analysis. This allows for reliable and expressive assessment of the safety of
  a system in early design stages.
\end{abstract}

\section{Introduction}
\label{sec:introduction}
\input{introduction}

\section{Case Study}
\label{sec:case-study}
\input{pandora}

\section{Probabilistic Failure Mode Modeling}
\label{sec:probabilistic-dcca}

\input{pandora-quant}

\section{Quantitative Safety Analysis}
\label{sec:quant-safety-analys}

\input{pandora-analysis}

\section{Related Work}
\label{sec:related-work}
\input{pandora-related}

\section{Conclusion and Future Work}
\label{sec:conclusion}

\input{conclusion}

\newpage

\bibliographystyle{eptcs}
\bibliography{cse-eigene,cse-andere}

\end{document}

%% file: introduction.tex
With rising complexity, larger machinery and bigger power consumption, more and
more systems become safety critical. At the same time, the amount of software
involved is growing rapidly, which increases the difficulty to build reliable
and safe systems. 

To counter this evolution, safety analysis has become a focus in many
engineering disciplines. Requirements for the development and life cycle of
safety-critical systems are now specified in many different norms like the
general IEC 61508~\cite{citeulike:3087012}, DO178-B~\cite{DO178-B} for aviation
or ISO 26262~\cite{ISO26262} for automotive. Although the standards address very
different application domains, they all require some sort of safety assessment
before a system is put into operation and require the use of formal methods for
systems in high risk areas. 

Safety assessments can be typically divided into two groups: qualitative and
quantitative assessments. Qualitative analysis methods like FTA (fault tree
analysis)~\cite{FThandbook02}, FMEA (failure modes and effects
analysis)~\cite{basicfmea} or HazOp (hazard and operability
analysis)~\cite{MOD5896} are used to determine causal relationships between
failure of individual components and system loss~\cite{IFAC05}.

These methods emerge from a long expertise in building safety-critical
systems. Their disadvantage is, that they mainly rely on skill and expertise of
the safety engineer. A potential safety risk will only be anticipated if the
engineer ``foresees'' it at design time. This becomes ever harder, because of
rising hardware and software complexity.

A new trend is to advance the analysis methods on a model-based level. This
means, that a model of the system under consideration as well as its environment
is built. The (safety) analysis is then not only grounded on the engineers skill
but also on the analysis of the model. In this way some causes for hazards can
be found much earlier. Errors found at early design stages are easier to remove
and redesign is less costly. Some examples of such methods are explained in
\cite{bretschneider-flap04, ISAACProject, IFAC05, bozzano03} which allow for
semi-automatic deduction of cause-consequence relationships between component
failures and loss of system.

However, qualitative analysis methods alone are not sufficient for norm
adherence (and for showing an adequate amount of safety). Another main criterion
for safety is the answer to the question: \emph{``What is/are the probabilities
  of any of the systems hazards?''} This is addressed by quantitative analysis.

Most quantitative approaches
either use the results of prior qualitative methods or a specific
(probabilistic) model of only failure effects and cascading without the
functional part of the system. This can lead to pessimistic estimations of the
actual hazard probabilities.
Newer methods like~\cite{COMPASS,GCW07,atr29} apply probabilistic model-checking
techniques to overcome this problem by the analysis of system models with both
failure and functional behavior. These methods use continuous time semantics for
the underlying stochastic models, which is well applicable for asynchronous
interleaved systems~\cite{Hermanns2000}. On the other hand, many safety critical
systems are developed using synchronous parallel discrete systems like clocked
bus systems and processing units.\footnote{For example SCADE Suite which is
  based on the synchronous data-flow language
  LUSTRE~\cite{halbwachs91synchronous} is widely used in industry for the
  development of safety critical systems, as it IEC 61508 and DO-178B certified
  code generators} Such synchronous parallel systems are better expressed in a
discrete time model~\cite{Hermanns2000}.

We propose a method for probabilistic model-based safety analysis for
synchronous parallel systems. We show how different types of failures, in
particular \emph{per-time} and \emph{per-demand}, can be accurately  modeled
and how such systems can be analyzed using probabilistic model checking.

The next section (Sect.~\ref{sec:case-study}) introduces a small case study and
gives a very short introduction to our qualitative model-based safety analysis
method. Sect.~\ref{sec:probabilistic-dcca} shows how accurate probabilistic
failure modeling can be achieved, Sect.~\ref{sec:quant-safety-analys} shows how
probabilistic models can accurately be analyzed. Sect.~\ref{sec:related-work}
gives an overview of related work in the field, and Sec.~\ref{sec:conclusion}
concludes the paper.


%% file: pandora.tex
For illustration purpose we use a small case study taken from
literature~\cite{DBLP:conf/safecomp/WalkerBP07}. Only a brief overview of
model-based (qualitative) safety analysis is given, more details can be found
in~\cite{IFAC05}. In short, the qualitative model-based safety analysis consists
of the following steps:

\begin{enumerate}
\item Formal Modeling of the functional system
\item Integration of the direct failure effects and failure automata forming the
  extended system model
\item Computation of all minimal critical sets using deductive cause consequence
  analysis (DCCA)
\end{enumerate}

\subsection{System Model of Case Study}
\label{sec:system-model-case}

\begin{figure}[h]
  \centering
  \includegraphics[width=0.6\textwidth]{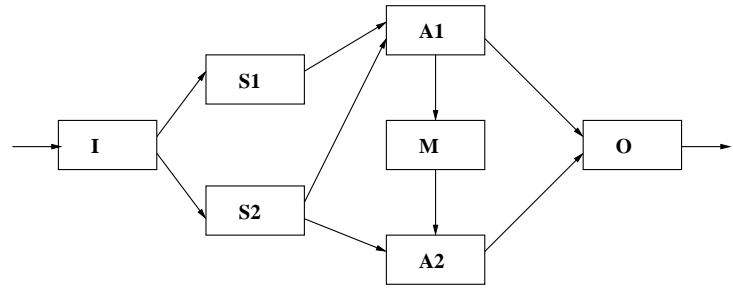}
  \caption{Backup Equipped System}
  \label{fig:backup-system}
\end{figure}

The case study consists of two redundant input sensors (S1 and S2) measuring
some input signal (I). This signal is then processed in an arithmetic unit to
generate the desired output signal. Two arithmetic units exist, a primary unit
(A1) and its backup unit (A2). The primary unit gets an input signal from both
input sensors, the backup unit only from one of the two sensors. The sensors
deliver a signal every 10ms. If the primary unit (A1) produces no output signal,
then a monitoring unit (M) switches to the backup unit (A2) for the generation
of the output signal. The backup unit should only produce outputs, when it has
been triggered by the monitor. The case study is depicted in
Fig.~\ref{fig:backup-system}.

\begin{figure}[h]
  \centering
  \psframebox[linewidth=1pt,framearc=.1]{
    \includegraphics[width=0.5\textwidth]{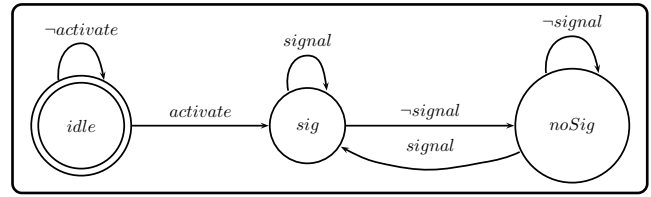}
  }
  \caption{Transition System for Second Arithmetic Unit}
  \label{fig:pandora-arith2}
\end{figure}

The system model is based on finite state machines which operate in a discrete
time synchronous parallel way. We use a graphical representation of the modules
of the actual model which is written for the SMV~\cite{SMV} model checker. The
initial state is marked with two circles, the transition labels are predicates.
The state of each module is exported as output so that it is visible in the
other modules. The modeling of the second arithmetic unit (A2) is shown in
Fig.~\ref{fig:pandora-arith2}.

Initially, the unit A2 is in state ``idle'' (i.e. a hot stand-by state where no
output is produced). It stays in this state until it gets activated (predicate
``activate'' is true) by the monitoring unit. It then is in state ``sig'', as
long as there is data available (predicate ``signal'' which means sensor
(S2) produces data). If there is no data available, the unit enters the state
``noSig'', as no signal can be produced. If the sensor starts delivering data
again, A2 switches back to state ``sig''. The other modules of the system are
modeled in a similar way, the system model is then the parallel composition of
all the modules.

\subsection{Formal Failure Modeling}
\label{sec:form-fail-model}

In this scenario, a variety of failures modes is possible. The sensors can omit
a signal (\emph{S1FailsSig}, \emph{S2FailsSig}), making it impossible for the
arithmetic unit to process the data correctly. The arithmetic units themselves
can omit producing output data (\emph{A2FailsSig}, \emph{A1FailsSig}). The
monitor can fail to detect that situation (\emph{MonitorFails}), either
switching if not necessary or not switching if necessary. The activation of the
A2 unit can fail (\emph{A2FailsActivate}). All these failure modes must now be
integrated into the (functional) model of the system.

The main idea to integrate failures correctly into a system is to separate
failure occurrence patterns and \emph{direct} failure effects. This allows for
conservative integration of failure modes, i.e. the original behavior of the
system is still possible if no failure occurs.

Occurrence patterns are modelled by \emph{failure automata}.  The most basic
failure automaton has two states, one state \emph{yes} modeling the presence of
the failure and a state \emph{no} modeling its absence. The transition
possibilities between the states determine the type of the failure mode. For
example: if the state \emph{yes} can never be left if it became active once, the
failure mode is called \emph{persistent}, if the state can non-deterministically
switch between \emph{yes} and \emph{no}, it is \emph{transient}. More complex
failure modeling can for example incorporate repairing or disappearance of the
failure after a given time interval.

\begin{figure}[h]
  \centering
  \psframebox[linewidth=1pt,framearc=.1]{
    \includegraphics[width=0.5\textwidth]{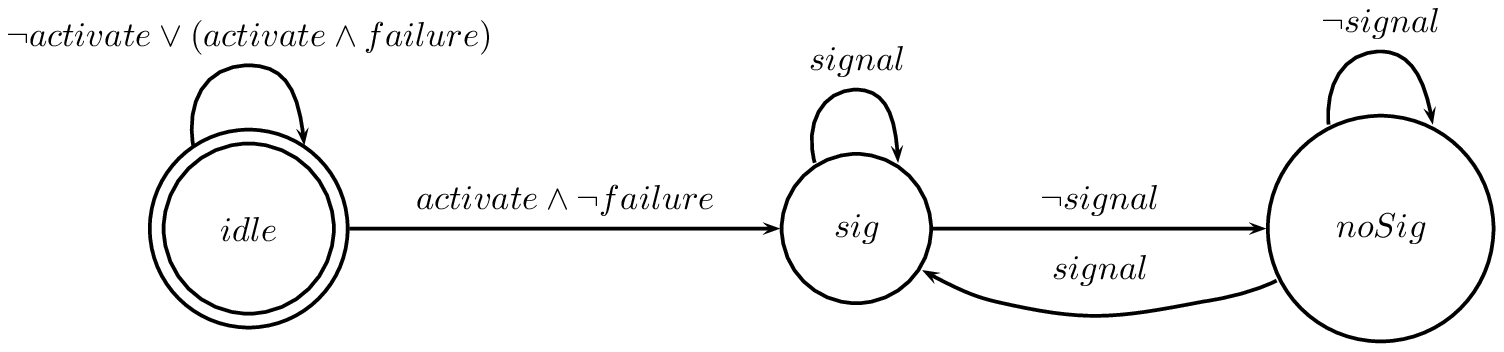}
  }
  \hspace{1cm}
  \psframebox[linewidth=1pt,framearc=.1]{
    \includegraphics[width=0.2\textwidth]{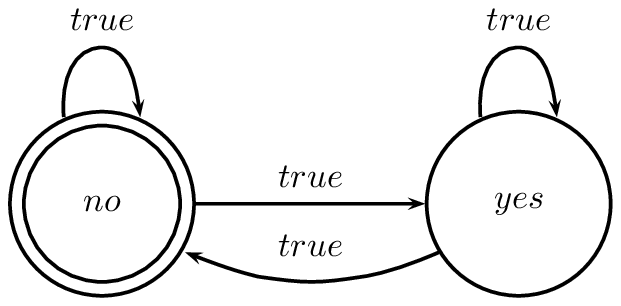}
  }
  \caption{Failure Mode \emph{A2FailsActivate}: Direct effect (left) and failure automaton (right)}
  \label{fig:failure-automata}
\end{figure}

The effects of the failure are modeled in the system model, using a predicate
\emph{failure} as the indication that a certain failure has appeared (the
corresponding failure automaton is in not in state ``no''). The formal system model
with the integration of the failure effects is called the \emph{extended system
  model}. The failure automata are then used in parallel composition with the
extended system model. The integration of the failure mode ``A2FailsActivate''
into the model can be seen on the left side in
Fig.~\ref{fig:failure-automata}. The extension of the functional model (as shown
in Fig.~\ref{fig:pandora-arith2}) has been straight forward: whenever the unit
A2 should be activated, it will only enter state ``sig'' (producing an output
signal) if and only if the corresponding failure automaton is in state ``no'',
i.e. the predicate $\neg failure$ holds.

The corresponding transient failure automaton is depicted on the right side of
Fig.~\ref{fig:failure-automata}. It is non-deterministic, as the transition
labels are always ``true'' which means that at any time it can stay in the
current state or leave. This models a randomly appearing and disappearing
(transient) failure mode. A more detailed explanation for general conservative
integration of failure modes can be found in
\cite{formal-failure-models-dcds07}.

\subsection{Qualitative Model-Based Safety Analysis}
\label{sec:dcca}

After the failures and their direct effects are integrated into the system
model, model-based safety analysis can be conducted on the extended system
model. This can be done with DCCA (deductive cause-consequence
analysis)~\cite{EDCC05, IFAC05}. DCCA provides a generalization of the FTA
minimal cut sets, called \emph{minimal critical sets} and is well suited to
synchronous contexts~\cite{DCCA-SCADE-SafeComp-07}. The generalization is proven
never to be worse than minimal cut sets and can often lead to a more accurate
analysis. It is based on the analysis of the whole system using temporal logics
and model-checking and does not rely on the manual construction of a FT. It can
be proven to be correct and complete~\cite{EDCC05} whereas FTA can be complete
but not correct and the minimal cut sets are too pessimistic (e.g. single point
of failures instead of necessary failure combinations). In order to conduct the
analysis, the synchronous parallel state machines are transformed to a Kripke
structure which is the used for model checking using the CTL branching time
temporal logic~\cite{clarke2000modelChecking}.

\begin{definition}{Critical Set / Minimal Critical Set}\\
  \label{def:formale-fmea}
  {\it    For a system SYS and a set of failure modes $\Delta$ a
    subset of component failures $\Gamma\subseteq\Delta$ is called critical for a system
    hazard, which is described by a predicate logic formula H if

    \[SYS \models{\bf E}(\overline{\Gamma}\until H)\;\;where\:\:\overline{\Gamma}
    := \bigwedge_{\delta\in(\Delta\setminus\Gamma)}\neg\:\delta\]
    
    $\Gamma$ is called a \emph{mcs (minimal critical set)} if $\Gamma$ is critical and no
    proper subset of $\Gamma$ is critical.}
\end{definition}

Informally the proof obligation means that: ``There exists (${\bf E}$) a run of
the system on which no failure mode of the set $\Delta\setminus\Gamma$ appears
($\overline{\Gamma}$) before ($\until$) the Hazard ($H$) appears.'' The failure
modes in the set $\Delta$ can appear in any order before the hazard.

The actual computation of the minimal critical sets starts with $\Delta =
\emptyset$, which equals functional correctness (testing if no occurrence of
failure modes is critical). Then single failures are checked and then the sets
are iteratively increased, omitting sets with critical subsets. Using this
approach only the \emph{minimal} critical sets are computed, as criticality as in
Def.~\ref{def:formale-fmea} is monotone. 

In the case study, using the SMV model checker~\cite{SMV} to compute the minimal
critical sets using the proof obligation of Def.~\ref{def:formale-fmea} leads to
the following resulting minimal critical sets for the hazard ``no signal at
output (O)'': Both arithmetic units fail (\{\emph{A2FailsSig},
\emph{A1FailsSig}\}), one arithmetic unit and the monitor fails
(\{\emph{A2FailsSig}, \emph{MonitorFails}\}) or (\{\emph{A1FailsSig},
\emph{MonitorFails}\}), the primary unit A1 and the second senor fail
(\{\emph{A1FailsSig}, \emph{S2FailsSig}\}), the monitor and the second sensor
fail (\{\emph{MonitorFails}, \emph{S2FailsSig}\}), both sensors fail
(\{\emph{S2FailsSig}, \emph{S1FailsSig}\}), the monitor fails and the activation
of A2 fails (\{\emph{MonitorFails},\emph{A2FailsActivate}\}) and the primary
unit fails and the activation of A2 fails (\{\emph{A1FailsSig}, \emph{A2FailsActivate}\})\footnote{On top of the minimal critical sets, temporal ordering information on the
  failure modes can also be automatically be extracted from the extended system model, in this
  case $\failOrder{MonitorFails}{A1FailsSig}$, i.e. the set \{\emph{A1FailsSig},
  \emph{MonitorFails}\} is only critical if the monitor fails \emph{before} the arithmetic
  unit, for details see \cite{ordered-dcca-2008}.}.


%% file: pandora-quant.tex
The qualitative analysis gives only insight into which combination of failures
can lead to a hazard. Nevertheless for an accurate estimation of the probability
that this happens, quantitative methods are required. Different types of
failures, particularly per-demand and per-time are possible which must be
accurately modeled. The overall approach for the proposed method for
probabilistic model-based safety analysis is an extension of the one in
Sect.~\ref{sec:case-study}:

\begin{enumerate}
\item Modeling: The same transition system as in the qualitative analysis can be used
\item[2a.] Failure Modeling: The same failure modes can be used
\item[2b.] Probabilistic Modeling: The modeling of the effects of the failure
  modes is augmented with \emph{failure probabilities} and \emph{failure rates},
  integration of the failure effect modeling is different depending on the type
  of failures.
\item[3.] Probabilistic model checking: The probability of the occurrence of the
  hazard $H$ is computed.
\end{enumerate}

\subsection{Temporal Resolution}
\label{sec:temporal-resolution}

For a realistic estimation of probability, the correct consideration of the
passing time is very important. In a discrete time context, there exists a basic
time unit which passes whenever the system performs a step (i.e. all parallel
finite state machines execute a transition). This is called the \emph{temporal
  resolution} $\tempres$ of the system. In synchronous parallel systems this
will usually be the basic clock of the system. This is the main difference
between CTMC and DTMC models. In DTMC if there is a probabilistic choice then
the system \emph{will} perform a step according to the probabilities
\emph{exactly} every $\tempres$ time units, whereas for continuous systems, it
is only possible to reason about the probability of reaching a state within a
given time $t$. For any possible time $t < \infty$ this probability is always
below 1. For DTMC models, the time $t$ is always an integral multiple of
$\tempres$ of the form $t=k \tempres, k\in \mathrm{N}$. Therefore for a clocked
system with synchronous parallel components (as opposed to interleaved
components), like many safety critical systems are, DTMC modeling is much better
suited than CTMC modeling~\cite{Hermanns2000}.

If asynchronous modeling is needed, e.g. vehicles are modeled which may either
move with a (normally distributed) velocity \emph{or} do not move at all, then
it must be done explicitly by specifying ``self-loops''. Making this explicit
allows for modeling of asynchronous behavior while conserving the temporal
semantics of the safety analysis technique.

\subsection{Failure Modeling for Quantitative Safety Analysis}
\label{sec:fail-model-quant}

To get accurate quantitative results, the correct probabilistic modeling of the
failure modes, especially of their occurrence pattern and their occurrence
probability is very important. At first, the type of the failure mode must be
determined. Two main types of probabilistic occurrence pattern for failure modes
exist. The first is \emph{per demand} failure probability, which is the
probability of the system component failing its function at a given demand
(comparable to \emph{low demand mode} in IEC 61508). The second is \emph{per
  time} probability, which is the \emph{rate} of failures over a given time
interval (comparable to \emph{high demand or continuous mode} in IEC 61508).

Which type of failure modeling is best fitting for a given failure mode can only
be decided on a case-by-case basis.  Sensor failures -- for example -- will very
often be modeled as a \emph{per time} failure mode, because sensors are often
active the whole time and are often modeled as a transient failure mode.  Other
failure modes, like the activation of a mechanical device, will often be modeled
as a \emph{per demand} failure, as a distinct moment of activation exists when 
the failure may occur.

The failure probability has an effect on the occurrence pattern of the failure
mode and is therefore reflected in the failure automata. For the probabilistic
modeling we use a graphical representation of the finite state machines used by
the probabilistic model checker PRISM~\cite{KNP02b}, which describes
\emph{probabilistic automata}. These are finite state machines like those used
for qualitative modeling (Sect.~\ref{sec:form-fail-model}, with the difference
that there is no non-determinism and the transitions are labeled with both an
activation condition and a transition probability.

Transitions are labeled with constraints of the form $(\phi;p)$ which means
that: ``If $\phi$ holds, then the transition is taken with probability
$p$''. Omitting $p$ means probability 1. A constraint of the form
$(\phi;p_1)\Or(\psi;p_2)$ means that the transition is taken with probability
$p_1$ if $\phi$ holds and with probability $p_2$ if $\psi$ holds. As the model
is deterministic $\phi \And \psi = \false$ always holds. As an additional
requirement in order to be a well defined DTMC model, for each state,
for each activation condition $\phi$, the outgoing transition probabilities must
always sum up to 1. This assures that the transition relation is total.

\subsubsection{Per-Time Failure Modeling}
\label{sec:per-time-failure}

\begin{figure}[h]
  \centering
  \psframebox[linewidth=1pt,framearc=.1]{
    \includegraphics[width=0.3\textwidth]{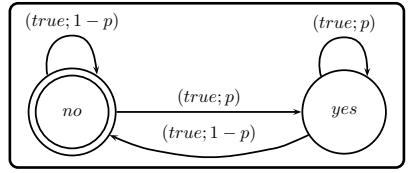}
  }
  \caption{Per-time Failure Automaton}
  \label{fig:per-time}
\end{figure}

A \emph{per time} failure can be modeled by adding the failure probability to
the transition from the state ``no'' to the state ``yes'' in a failure automaton
as shown in Fig.~\ref{fig:per-time}, as it can occur at any time of the system
run. So basically the non-deterministic transitions of the transient failure
automaton are replaced by probabilistic transitions.

The probability $p$ is computed from the given per-time failure rate $\lambda$
and the temporal resolution $\tempres$ in order to approximate the continuous
exponential distribution with parameter $\lambda$ in the discrete time
context. This distribution function is shown in Eq.~\ref{eq:exp-dens}. It
computes the probability that the failure occurs before time $t$ (i.e. the
random variable $X$ has a value less than or equal $t$). This distribution is
often used for per-time failure modeling, especially in continuous time models,
therefore it is important to be expressible in the discrete time context.

\begin{equation}
  \label{eq:exp-dens}
  P(X\leq t) = \int_0^t e^{-\lambda t} = 1 - e^{-\lambda t}
\end{equation}

In the discrete time model of DTMCs, the occurrence probability of the failure
mode before time $t = k \tempres$ ($k$ time steps of length \tempres) as modeled
with the failure automaton (Fig.~\ref{fig:per-time}) forms a geometric
distribution and is shown in Eq.~(\ref{eq:prob-bernoulli}). As $p$ is the
per-demand occurrence probability, $1-p$ is the probability that the failure
does not occur and $P(X > k)$ is the probability that the failure does not occur
for $k$ time steps.

\begin{equation}
  \label{eq:prob-bernoulli}
  P(X \leq k) = 1 - P(X > k) = 1 - (1 - p)^k
\end{equation}

Using the identity $e^x = lim_{n \rightarrow \infty}(1 + \frac{x}{n})^n$ the
continuous exponential distribution
can be approximated with the discrete geometric distribution as shown in
Eq.~(\ref{eq:prob-approx}). For longer time intervals $k$ approximates $n$.
Then $\lambda \tempres$ can be substituted as probability $p$ in the per-time
failure automaton (Fig.~\ref{fig:per-time}) and in Eq.~\ref{eq:prob-bernoulli}.

\begin{equation}
  \label{eq:prob-approx}
  1 - e^{-\lambda t} = 1 - lim_{n \rightarrow \infty}
  \left(1 + \frac{-\lambda t}{n}\right)^n = 1 -
  lim_{n \rightarrow \infty}
  \left(1 - \frac{\lambda k \tempres}{n}\right)^n \approx 1 - (1 -
  \lambda\tempres)^k
\end{equation}

\begin{figure}[h]
  \centering
  \includegraphics[width=0.49\textwidth]{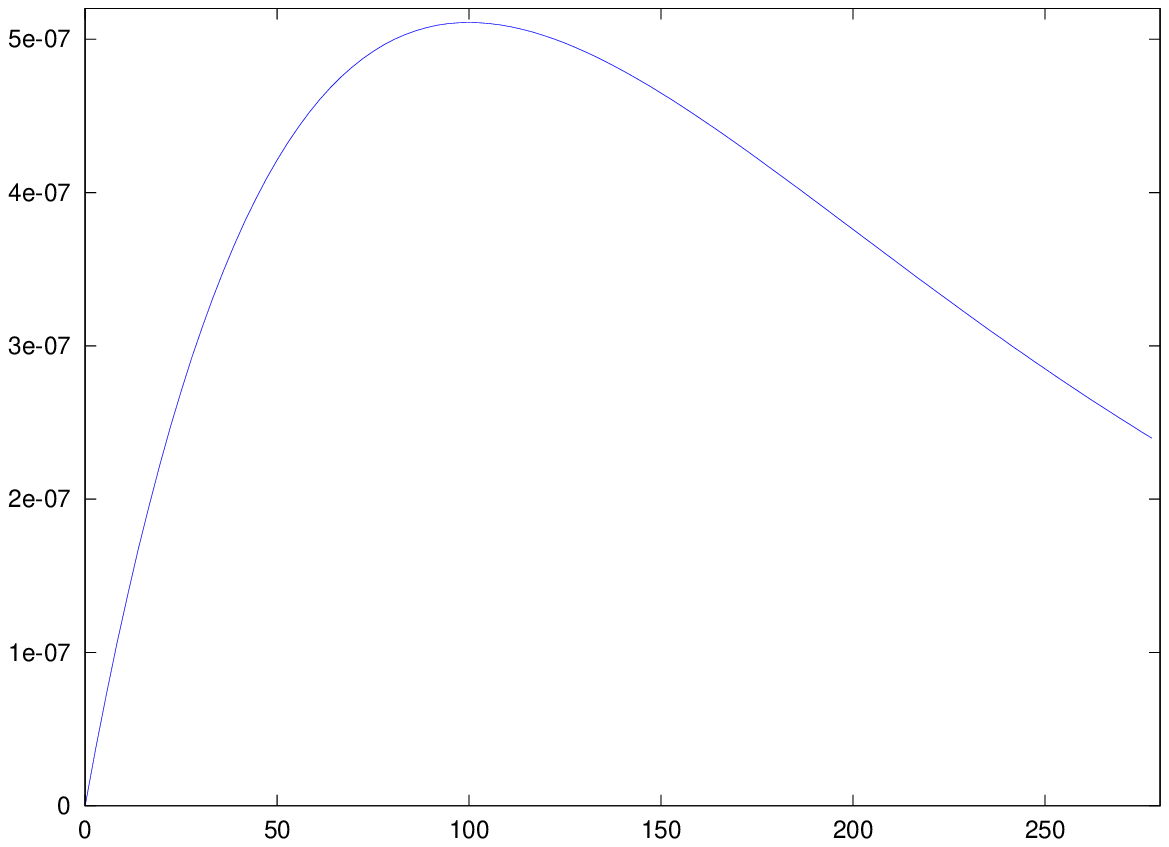}
  \includegraphics[width=0.49\textwidth]{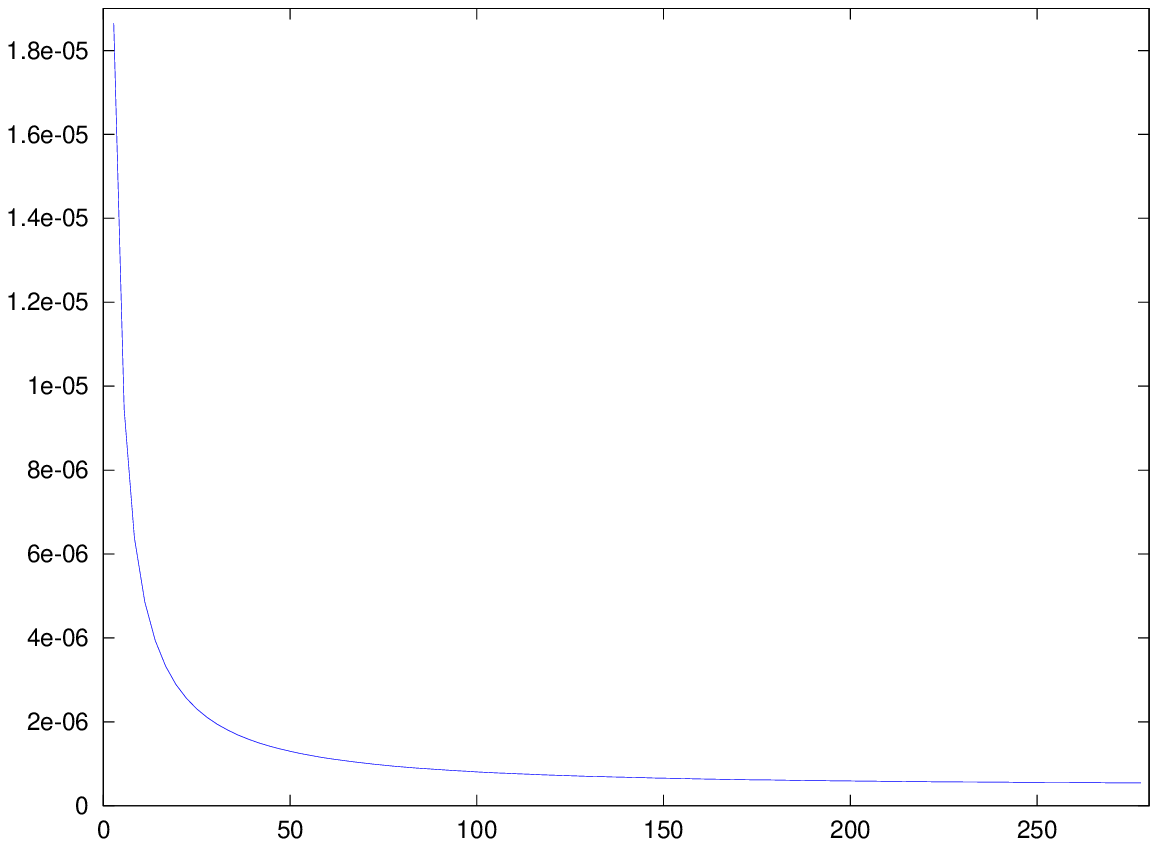}
  \caption{Absolute and relative error for discrete approximation}
  \label{fig:approx-error}
\end{figure}

The left graph in Fig.~\ref{fig:approx-error} shows the absolute approximation
error $\epsilon(t)=|(1 - e^{-\lambda t}) - (1 - (1 - \lambda \tempres)^k)|$ for
$\lambda = 10^{-2}\frac{1}{h}$, $\tempres = 1s$ and $t = k \tempres$. The
maximum is reached at $t = \frac{1}{\lambda}$ with a function value for the
exponential distribution $1 - e^{\frac{-\lambda}{\lambda}} = 1 - \frac{1}{e}
\approx 0.63212$. The approximation error at $t=\frac{1}{\lambda}$ is
approximately $5.1095\cdot 10^{-7}$ which is several orders of magnitude lower
than the function value. This can be seen on the right graph of
Fig.~\ref{fig:approx-error} which shows the relative error
$|\frac{\epsilon(t)}{1-e^{-\lambda t}}|$. In both figures the x-Axis is the time
axis $t$ in hours. The approximation error decreases as $t$ and $k$ increase
(the relative error after $t=\frac{1}{\lambda}$).

\subsubsection{Per-demand Failure Modeling}
\label{sec:per-demand-failure}

The correct modeling of a \emph{per demand} failure mode is more complex. Not
only the occurrence pattern, but also the modeling of the direct effect must be
adapted to probabilistic modeling. The challenge is that the system which
executes a demand and the failure automaton which indicates whether the demand
succeeds or fails must take a step at the same time and the failure automaton is
not allowed to take a transition if there is no demand, or else the computed
probabilities are not correct.

\begin{figure}[h]
  \centering
  \psframebox[linewidth=1pt,framearc=.1]{
    \includegraphics[width=0.5\textwidth]{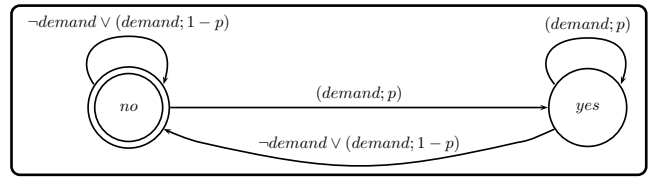}
  }
  \caption{Per-demand failure automaton}
  \label{fig:per-demand-failure-automaton}
\end{figure}

For the occurrence pattern, a predicate \emph{demand} is defined which indicates
that there is a demand to the safety critical component. The per-demand failure
automaton can only leave the state ``no'' if \emph{demand} holds, see
Fig.~\ref{fig:per-demand-failure-automaton}. The state ``no'' has two
transitions, one loop which is active if there is either no demand
$\neg$~\emph{demand} or if there is a demand but the demand succeeds with
probability $(1-p)$ ($p$ being the per-demand failure probability). The state
``no'' is only left with probability $p$ if \emph{demand} holds.

The integration of the direct effect of a per-demand failure into the system
model is a bit more complicated. It is illustrated via the integration of the
failure mode of the second arithmetic unit \emph{A2FailsActivate} of the example
in Sect.~\ref{sec:form-fail-model}. At first, the predicate $activate$ which
means the hot stand-by mode must be left, is used to define the demand as
$demand := activate$, i.e. the demand is the activation command for the
secondary arithmetic unit. The activation can only fail if it is actually
called, therefore it is modeled as a per-demand failure.

Now if the $demand$ holds, the failure automaton and the system can make a
transition in parallel. Therefore the information whether the demand has been
met is available \emph{after} the transition has been taken. In order to get the
timing correct, an additional state is introduced which represents either
failure or success of the demand, depending on the state of the failure
automaton.

\begin{figure}[h]
  \centering
  \psframebox[linewidth=1pt,framearc=.1]{
    \includegraphics[width=0.7\textwidth]{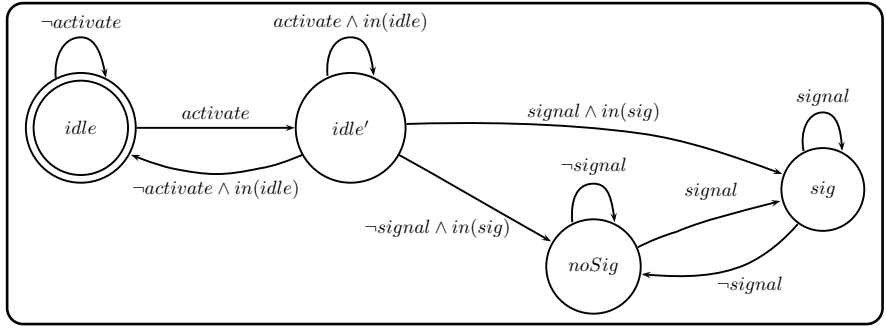}
  }
  \caption{Transformation for \emph{per demand} failure mode}
  \label{fig:per-demand-trans-failure}
\end{figure}

For the example this can be seen in Fig.~\ref{fig:per-demand-trans-failure}. The
\emph{demand} ($activate$) is possible in state $idle$. The original successor
states of $idle$ were $sig$ if the demand was successful and $idle$ if the
demand failed (see Fig.~\ref{fig:failure-automata}). For the probabilistic
modeling the state $idle'$ is added to represent a failed demand if at the same
time the failure automaton is in state ``yes'' and a successful demand if the
failure automaton is in state ``no''. To preserve the original behavior of the
system, the transitions of $idle$ and $sig$ must be added to the state $idle'$
in conjunction with the predicates $in(idle)$ or $in(sig)$ which are defined as shown
in Eq.~(\ref{eq:in-preds1}) and (\ref{eq:in-preds2}).

\begin{eqnarray}
  \label{eq:in-preds1}
  in(idle) & := & state = idle \vee (state = idle' \wedge failure = yes) \\
  \label{eq:in-preds2}
  in(sig) & := & state = sig \vee (state = idle' \wedge failure = no)
\end{eqnarray}

These predicates must then also be exported from the modules and be substituted
in all other places where $state = idle$ or $state = sig$ appears (predicates
for transitions, logic formulas etc.). If this is done, the observable behavior
of the automaton is the same as the original failure effects modeling as defined
in~\cite{formal-failure-models-dcds07} and shown in
Fig.~\ref{fig:failure-automata} for the example, but with a correct modeling of
the per-demand failure occurrence pattern and direct effect.

\subsubsection{General Per-Demand failure integration}
\label{sec:general-per-demand}

The last section illustrated the integration of a per-demand failure mode for
the given example. In general, this integration requires first to define what
constitutes a ``demand'' for a given failure mode. This typically results in a
state formula, which divides the set of all states (of the system) in states
where there is a demand for the functionality of the component and those where
there is no demand. Now, for all states $s\in S$ where a demand to a component
is exists, let $A_s \subseteq S$ be the set of direct successors states if there
was a successful demand and $B_s \subseteq S$ if it was unsuccessful, i.e.

\begin{quote}
  $A_s := \{a \in S | \exists$ transition from $s$ to $a$ with activation
  condition $demand \And failure = no$\}
\end{quote}
\begin{quote}
  $B_s := \{b \in S | \exists$ transition from $s$ to $b$ with activation
  condition $demand \And \neg failure = no$\}
\end{quote}

There can also be successor states of $s$ that are not in $A_s \cup B_s$, the
successors if there was no demand (where $demand$ as in the definition of $A_s$
and $B_s$ does not hold) to the safety critical system. These states do not play
any role in the following construction, but are one of the main causes that the
construction is required. The reason is that if there is no demand and therefore
the successor of $s$ is not in $A_s \cup B_s$, the failure automaton is not
allowed to make a transition.

Let $T_{A_s} := \{$ Transitions from $s$ to $d$ with $d\in A$\}, $\Phi_{A_s} :=
\{\phi | \; \phi \And failure = no$ is activation condition of any $t \in
T_A\}$, and define $T_{B_s}$ and $\Phi_{B_s}$ analogously, let $T := T_{A_s}
\cup T_{B_s}$, $\Phi := \Phi_{A_s} \cup \Phi_{B_s}$.  As the model is required
to be deterministic, $\forall \phi_i,\phi_j \in \Phi_A, i\neq j: \phi_i \And
\phi_j = \false$ holds, the same holds for $\Phi_B$. Now \emph{demand} can be
defined as $(state = s) \And \disjunction_{\phi \in \Phi} \phi$.

To integrate a \emph{per demand} failure, the following steps must be executed
for each automaton $M$ in which direct effects of a failure mode are modeled,
resulting in an automaton $M'$:

\begin{enumerate}
\item Define $M'$ with the same states as $M$. All transitions but those
  starting from state $s$ and going to states in $A \cup B$ are kept.
\item Define a new state $s'$ in $M'$ which represents being in any state $d \in
  A_s \cup B_s$, add a transition from $s$ to $s'$ where the activation
  condition is \emph{demand}
\item Define a \emph{decide} automaton which is used to specify which state
  $d\in A_S \cup B_s$, the state machine $M'$ really has, the set of possible
  states for the decision automaton is $A_s\times B_s\cup \{undef\}$.
\item Add failure probabilities to the corresponding failure automaton as in
  Fig.~\ref{fig:per-time}.
\item For each transition from $d \in A_s\cup B_s$ to a successor $d'$ with
  activation condition $\psi$ add a transition from $s'$ to $d'$ in $M'$ with
  the activation condition $\psi \And in(d)$
\item In the \emph{decide} automaton, let $undef$ be the initial state and add a
  to transition to a state $d = [a, b] \in A_s\times B_s$ if the transition from
  $s$ to $a$ is activated if $failure = no$ holds and the transition from $s$ to
  $b$ is activated if $\neg failure = no$ holds.
\item Add the corresponding transitions also between the states $d \in A\times
  B$ of the \emph{decide} automaton if a \emph{demand} is possible in two
  consecutive time steps
\end{enumerate}

This construction allows the failure automaton to make a transition from
\emph{no} iff \emph{demand} holds, $M'$ makes a transition from $s$ to $s'$ and
the respective \emph{decide} automaton makes a transition to one of its
states. For each of the original successor states $d\in A_s\cup B_s$, the
predicate $M' = d$ is not enough to characterize the fact that $M'$ is in state
$d$. Therefore the predicate $in(d)$ as shown in equation (\ref{eq:in-d-A}) and
(\ref{eq:in-d-B}) is introduced and replaces the occurrences of $M' = d, d \in
A_s \cup B_s$, for $d\in A_s$ as shown in Eq.~(\ref{eq:in-d-A}) and for $d \in
B_s$ as shown in Eq.~(\ref{eq:in-d-B}).

\begin{eqnarray}
  \label{eq:in-d-A}
  in(d) & := & M' = d \Or (M' = s' \And  decide = [s, d] \And  failure = no)  \\
  \label{eq:in-d-B}
  in(d) & :=  & M' = d \Or  (M' = s' \And  decide = [s, d] \And  \neg failure = no) 
\end{eqnarray}

Because of the determinism and the requirements of a total transition relation,
the successor state of $s$ is always well-defined, as for each label predicate
$l\in \mathcal{P}($atomic predicates of $\phi \in \Phi_A \cup \Phi_B$) there is
exactly one transition to a state if $failure = no$ and exactly one transition
if $\neg failure = no$.

In the example of Sect.~\ref{sec:per-demand-failure} the decide automaton has
been omitted in order to illustrate the idea of per-demand failure mode
integration without too much complexity. As $idle'$ only represents two
different states, and therefore the distinction using the $failure$ predicate is
enough. But for more complex systems with several possible per-demand failures,
the general construction using the additional decide automaton is necessary.


%% file: pandora-analysis.tex
The probabilistic models of Sect.~\ref{sec:probabilistic-dcca} can now be used
to compute the overall hazard probabilities of the system using probabilistic
logics and probabilistic model checkers.

\subsection{Probabilistic Model Checking}
\label{sec:prob-model-check}

For discrete time probabilistic models, the probabilistic logic probabilistic
computational tree logic (PCTL) is used.  For continuous time model, continuous
stochastic logic (CSL) is used. Their respective semantics can be found
in~\cite{KNP07a,Hansson94alogic}.

PCTL is a probabilistic variant of CTL. Instead of Kripke structures as in CTL
model-checking, labeled Markov chains are the system model for PCTL model
checking. Both are very similar, the main difference is that in Kripke
structures there is either a transition between states or there is none. In
labeled Markov chains there is either no transition between states or it has an
assigned probability. PCTL formulas are of the form:

\[P_{\sim p}[\phi] \sim \in \{<,\leq,>,\geq\}\]

and assert that for the probability $P$ of the path formula $\phi$ the equation
$P \sim p$ holds. This means (for example), for a given system the given formula
$\phi$ holds with a probability $P$ and $P \leq p$.  Intuitively this means,
that the probability that $\phi$ holds on a system trace starting in the initial
state is less than or equal to $p$. The other operators are defined analogously.

It is also possible to (automatically) calculate the probability directly. The
according syntactic expression is: $P_{=?}[\phi]$. Instead of giving a result
\emph{true} or \emph{false}, this returns the actual probability that the
formula $\phi$ holds on a given system trace.

The path operator available in PCTL is the \until operator. ``Globally (G)'' and
``Eventually (F)'' operators can be derived from this operator ($G \phi := \neg
(\true \until \neg \phi)$ and $F \phi := (\true \until \phi )$). Negation of a formula is
formulated by using $P_{\leq 0}$. In order to compute the probability of the
negation of a formula $\phi$, $1 - P_{=?}[\phi]$ can be used. The existence of a
trace on which $\phi$ holds can be checked with $P_{>0}[\phi]$.

The exact definitions of semantics and probability measures can be found in
\cite{KNP07a,Hansson94alogic}. Algorithms to check PCTL formulas can be found in
\cite{KNP07a}. Efficient model checking algorithms are implemented in tools like
PRISM~\cite{KNP02b} or MRMC~\cite{KatoenKZ_QEST05}. A comparison of the
efficiency and memory usage of different probabilistic model checkers including
stochastic simulation based model checkers using hypothesis testing can be found
in \cite{JansenKOSZ_HVC07}.

\subsection{Computation of Hazard Probabilities}
\label{sec:fail-prob}

PCTL proof obligations can then be used together with the probabilistic model to
compute the precise probability that the hazard occurs, given the per-time failure
rates and the per-demand failure probabilities of the failure modes.

The semantics for PCTL are defined for infinite system traces. This cannot be
used to compute the hazard probability, as it would always be either 1 if the
hazard can occur or 0 if the hazard cannot occur.

Instead, the analysis is conducted for a given time range. What is computed is
the probability that the hazard occurs in $k$ time steps of length \tempres. To
accomplish this, the \emph{bounded until} temporal operator is used. If $\phi
\until^{\leq k} \psi$ holds, then there exists a bound $j \leq k$, so that
$\psi$ becomes \emph{true} after no more that $j$ steps and $\phi$ is
\emph{true} for all time steps $i < j$.  Using the bounded until operator, the
occurrence probability of the hazard $H$ in time $t = k \tempres$ is computed
via Eq.~(\ref{eq:pctl-prob-hazard}) which is basically the bounded PCTL DCCA
formula of Def.~\ref{def:formale-fmea} for an empty minimal critical set.

\begin{equation}
  \label{eq:pctl-prob-hazard}
  P_k(H) := P_{=?}[true \until^{\leq k} H]   
\end{equation}

It is important to note, that the direct analogon to the DCCA formula
$\left(P_k(\Gamma) := P_{=?}[\overline{\Gamma} \until^{\leq k} H], \Gamma \neq
\emptyset \right)$ does not compute the probability of the failure modes in the
minimal critical set $\Gamma$ causing the hazard $H$. The reason is that
$\overline{\Gamma} \until^{\leq k} H$ limits the set of traces to those where
\emph{only} those failure modes in $\Gamma$ can appear. This is adequate to
analyze whether these failure modes are sufficient to cause $H$ as it is done
with DCCA (which is a \emph{worst case} analysis).

Nevertheless for probabilistic analysis the probability of \emph{all} traces
where $\Gamma$ is the cause of $H$ is important, i.e. also the traces where
$\Gamma$ causes $H$ but other failure modes appear but without any direct effect
on the occurrence of $H$. For example, if A2 fails transiently before the
monitor fails and then A1 fails, the cause for ``no signal output'' is the
minimal critical set \{\emph{A1FailsSig}, \emph{MonitorFails}\}. But the
probability of the trace where \emph{A2FailsSig} appears would not be
considered, as $\overline{\Gamma} \until^{\leq k} H$ limits the traces to those
where only \emph{A1FailsSig} and \emph{MonitorFails} appear.

Instead of pure DTMC models as shown here, it is also possible to use MDPs
(Markov decision processes) which allow a non-deterministic choice between
different distribution functions in a state. Instead of exact probabilities, MDP
analysis allows for computation of minimal and maximal probabilities. For safety
analysis, the \emph{maximal} (worst case) probability of the hazard is of
interest, as shown in Eq.~\ref{eq:pctl-prob-hazard-mdp}.

\begin{equation}
  \label{eq:pctl-prob-hazard-mdp}
  Pmax_k(H) := Pmax_{=?}[true \until^{\leq k} H]   
\end{equation}

\subsection{Quantitative Results}
\label{sec:prob-results-pandora}

For the analysis of the example case study, all failure modes were modeled as
\emph{per time} failure modes with the exception of the \emph{per demand}
failure \emph{A2FailsActivate}. The minimal critical sets are given in
Sect.~\ref{sec:dcca}.

Assuming an error rate of $10^{-2}\frac{1}{h}$ for each \emph{per time} failure
mode and using a temporal resolution $\tempres = 10ms$, this translates to a
per-step failure probability of $p_{fail} = 2.\overline{7} \cdot 10^{-8}$ for
the per-time failure modes. For the failure mode ``A2ActivateFails'', a
\emph{per demand} failure probability of $10^{-4}$ is assumed.

Using PRISM the computation of the hazard ``no output delivered'' can be
computed for a for $k = 360000$ (i.e. running time of 1h) using the
probabilistic modeling as described in Sect.~\ref{sec:probabilistic-dcca} The
result is shown in Eq.~\ref{eq:p-quant-pandora}. The analysis was conducted on
an AMD64 3Ghz and needed ca. 140s to complete. In this case the dominant factor
in the analysis time is the large number of iterations, as $k$ matrix
multiplications are required. This can very likely be made much faster using new
approaches for probabilistic model
checking~\cite{BES09,DBLP:conf/qest/BarnatBCCT08}.

\begin{equation}
  \label{eq:p-quant-pandora}
  P_{Quant}(H) =  2.964375\cdot 10^{-17}
\end{equation}

If a quantitative method based on the a-posteriori analysis of the qualitative
results like FTA is used, the estimation of the actual hazard probability is
either very pessimistic (if the dependency of the ordering is not considered) or
it can get very complex. For example to increase the accuracy, the system model
must be analyzed further and the dependencies of the effects of the various
failure modes must be explored (the model is based on finite state machines and
time passes if a state changes which can then be detected in the next time
step). In addition, as many as possible of the possible occurrence patterns of
the failure modes must be enumerated to enhance the result. On the other hand,
the probabilistic analysis can do this automatically and is much less error
prone than such an a-posteriori analysis of the model.


%% file: pandora-related.tex
Several approaches for quantitative safety analysis exist which rely mainly on
the analysis of a previous qualitative analysis. This can be either fault tree
analysis (FTA)~\cite{FThandbook02} or methods to compute the critical failure
combinations directly from a model. One example is DCCA~\cite{IFAC05}, other
methods relying on fault injection were developed in the ESACS
project~\cite{bozzano03:esacs} like the FSAP/NuSMV-SA framework~\cite{bozzano03}
or in the ISAAC project~\cite{ISAACProject} were SCADE was used for modeling and
safety analysis~\cite{akerlund04-scade}. In order for the estimation of the
global hazard probability, the ``FTA-formula'' is used (see
Eq.~(\ref{eq:fta-eq})) which gives an upper bound for the hazard probability.

\begin{equation}
  \label{eq:fta-eq}
  P(H) \leq \ftaFormula P(\delta)
\end{equation}

It is clear, that no ordering information is considered in this formula as it
depends on all probabilities being independent, so the estimation can be very
pessimistic. With dynamic fault trees (DFT)~\cite{cepin02}, ordering information
and dependencies can be analyzed, they can partly be deduced from the models
directly~\cite{bozzano03a,ordered-dcca-2008} and can give more accurate results
than FTA. Nevertheless it is still based on a-posteriori analysis of qualitative
results with the corresponding disadvantages.

In the COMPASS project~\cite{COMPASS} the FSAP-NuSMV/SA~\cite{bozzano03}
framework is combined with the MRMC~\cite{eemcs1545} probabilistic model checker
to allow for the analysis of systems for Aerospace applications specified in the
SLIM~\cite{slim-bozzano-katoen} language, which is inspired by
AADL~\cite{aadl-architecture,aadl-error-annex}. The hybrid behavior of the SLIM
models and all internal transitions are removed by lumping and the resulting
interactive Markov chain is analyzed with MRMC.

Another approach for probabilistic safety analysis is probabilistic failure
modes and effects analysis (FMEA)~\cite{GCW07} where per-time failure behavior
is integrated into system models via failure injection. After then, FMEA tables
are computed which are often used in industrial safety analysis processes.

Both these approaches use continuous Markov chains models (CTMC) and only use
per-time failure mode modeling with failure rates. The semantically correct
integration of per-demand failure mode into continuous time models is difficult
at best and in general not possible (non-deterministic behavior may occur). But
it would be very interesting to see how these methods, especially the SLIM
modeling language can be used for synchronous parallel system and in discrete
time models.

In the AVACS project~\cite{avacs03}, another model-based safety analysis
approach based on statecharts was developed~\cite{atr29}. It also uses
continuous time models, but an additional difference is, that the hazard
probability is analyzed for each minimal cut set. As explained in
Sect.~\ref{sec:fail-prob}, the probability for a given set to cause the hazard
is not easy to compute. Therefore it is not done in~\cite{atr29}, but an
``importance analysis'' is conducted, analyzing single critical combinations and
rating them according to their contribution to the global hazard probability.


%% file: conclusion.tex
We showed how model-based safety analysis can be conducted on extended system
models with probabilistic failure behavior. Per-demand as well as per-time
failure modes can be integrated and the overall probability of a hazard
occurrence in a given time interval can be computed. This quantitative analysis
is well applicable to synchronous parallel systems.

The method has been applied to much larger case studies with more than $10^8$
states and due to recent developments in the area of probabilistic model
checking, e.g. using the GPU~\cite{BES09}, parallel model
checking~\cite{DBLP:conf/qest/BarnatBCCT08}, abstraction
techniques~\cite{DBLP:conf/tacas/KatoenKZJ07,BaierDArgenioGroesser2005:entcs,1427803}
and also simulation based stochastic hypothesis testing~\cite{Younes05s} still
larger systems will likely be analyzable in the future. Hypothesis testing is
very interesting, as it allows to give probabilistic results with confidence
intervals based on several simulation runs. Using it, probabilistic analysis can
give reliable results even if complete model checking is not possible.

In any way, this model-based analysis can give insights into the properties of a
safety critical system in the in early phases of the design, where the cost of
redesign is lowest.

At the moment, a framework is developed which allows for both qualitative and
quantitative analysis on the same model, automating the complex integration of
per-demand failures as presented here. The given approximation of the continuous
distribution and the time discretization worked well in the considered case
studies. Nevertheless additional estimation of the accuracy of the results is
aspired, depending on the temporal resolution and the discrete probabilities,
especially for probabilistically modeled environment behavior other than failure
modes. Very beneficial would be the direct integration into tools like
Topcased~\cite{Vernadat2006} or SCADE, which requires additional work for the
transformation from these into our framework. But the semantic proximity of the
SCADE execution models and the already accomplished integration of DCCA into
it~\cite{DCCA-SCADE-SafeComp-07} are evidence that this is possible.
